\documentstyle[epsfig,longtable]{aipproc}

\begin{document}

\title{Spectroscopy of Quasar Candidates from 
SDSS Commissioning Data}
 
\author{Xiaohui Fan$^1$, Michael A. Strauss$^1$, James Annis$^4$,
James E. Gunn$^1$, Gregory S. Hennessy$^2$, Zeljko Ivezic$^1$, 
Gillian R. Knapp$^1$, Robert H. Lupton$^1$, Jeffrey A. Munn$^3$, Heidi J. Newberg$^4$,
Donald P. Schneider$^5$, and Brian Yanny$^4$
for the SDSS Collaboration}
\address{(1) Princeton University Observatory, Princeton, NJ 08544\\
(2) U.S. Naval Observatory, 3450 Massachusetts Ave., NW, Washington,
DC 20392\\
(3) US Naval Observatory, Flagstaff Station, PO Box 1149, Flagstaff,
AZ 86002\\
(4) Fermi National Accelerator Laboratory, PO Box 500, Batavia, IL 60510 \\
(5) Astronomy \& Astrophysics, 525 Davey Lab, University Park, PA 16802}
\maketitle
\begin{abstract}
The Sloan Digital Sky Survey has obtained images in five broad-band
colors for several hundred square degrees.  We present color-color
diagrams for stellar objects, and demonstrate that quasars are easily
distinguished from stars by their distinctive colors.  Follow-up
spectroscopy in less than ten nights of telescope time has yielded 22
new quasars, 9 of them at $z> 3.65$, and one with $z = 4.75$, the
second highest-redshift quasar yet known.  Roughly 80\% of the high-redshift
quasar candidates selected by color indeed turn out to be
high-redshift quasars. 
\end{abstract}


The Sloan Digital Sky Survey (SDSS; \cite{GunnWeinberg,SDSS}) will use a dedicated 2.5m
telescope at Apache Point Observatory in Southeast New Mexico
to obtain CCD images to $ \sim 23^m$ in five bands ($u', g', r', i',
z'$; \cite{Fukugita}) over 10,000 square degrees of high Galactic
latitude sky.  The imaging camera (\cite{CameraPaper}) contains 30 $2048
\times 2048$ and 24 $2048 \times 400$ CCDs in its focal plane, and
takes data at a rate of 20 square degrees an hour in drift-scan mode
in all five colors; the data rate is roughly 1 Gbyte per square
degree.  Specialized software has been written to carry out
astrometric and photometric calibration of the data, and to find and
measure the properties of all objects detected in the images.  The
brightest $10^6$ galaxies and $1.5 \times 10^5$ quasar candidates 
will be followed up spectroscopically on the
same telescope, using a pair of double spectrographs fed by a total of
640 fibers. 

The SDSS obtained first light in imaging mode in May 1998, and is now
undergoing intensive commissioning.  We report here on the
distribution of stellar objects in color-color space, the selection of
quasar candidates, and follow-up spectroscopy with the Apache Point
3.5m telescope. 


The SDSS will use the colors and morphology of objects to identify
quasar candidates from the photometric data: objects with stellar
appearance and colors that lie well outside the stellar locus in
color space will be flagged for spectroscopic investigation.


\begin{figure}
\centerline{\epsfig{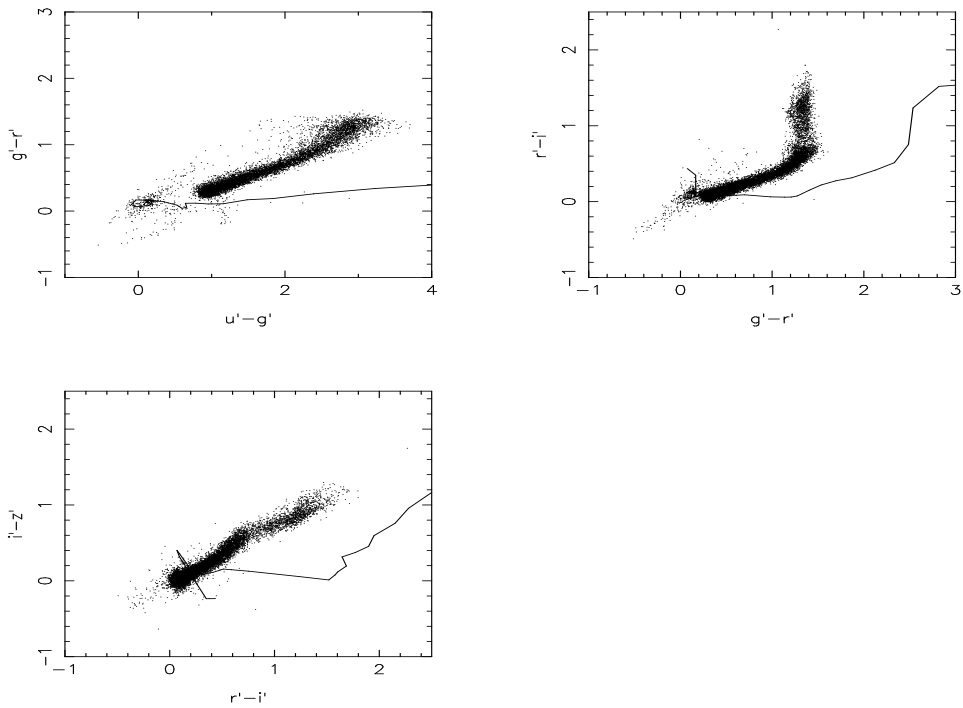}}
\vspace{10pt}
\caption{
Simulated distribution of stellar objects in projections of SDSS color
space over 10 square degrees towards the North Galactic Pole, which includes
quasars, normal stars, white dwarfs and compact
emission line galaxies to $r' = 20$.  The solid line is the mean locus of
quasars as a function of color. }\label{colorsim}
\end{figure}

The distinction between quasars and stars in color-color space is
illustrated in Figure~\ref{colorsim}, which shows model distributions
of 
stars and quasars in a series of three SDSS color-color diagrams, from
the simulations of ref.~\cite{Fan}.  These simulations put in realistic
SED's for stars, quasars, and compact emission-line galaxies, and
attempt to model the stellar populations and spatial distributions of
stars for the North Galactic Pole.  The mean
locus of quasars as a function of redshift is shown as the solid line;
for $z < 2.5$, quasars are very blue in $u' - g'$, and can be
distinguished quite easily from stars (and hot white dwarfs as well, which tend
to be bluer in $g'-r'$; see the discussion in ref.~\cite{Fan}).  At higher
redshifts, the Lyman forest, and eventually, Lyman-limit systems, move
through the SDSS filters, causing the colors to become 
redder.   Note that at most redshifts, the quasar locus is
well-separated from the stellar locus; the pernicious exception is
quasars at $z \approx 2.8$, which have very similar broad-band colors
to an F star.  The reddest bands will permit identification of
quasars with redshifts higher than six. 


\begin{figure}
\centerline{\epsfig{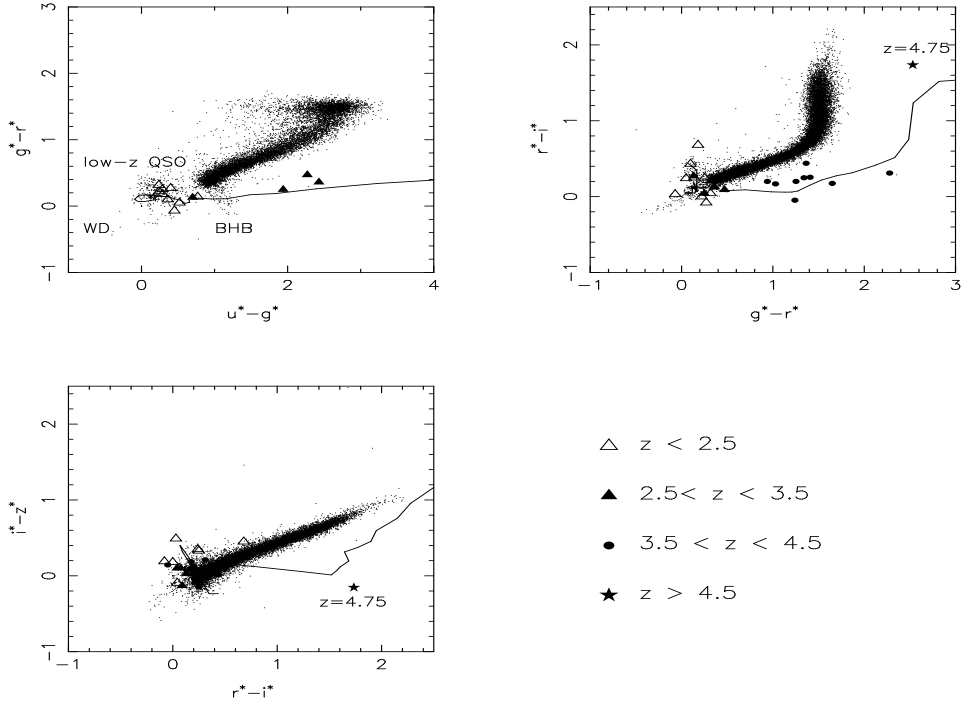}}
\vspace{10pt}
\caption{
Observed color-color diagrams of
20 square degrees from the SDSS test data ($r^*<20$).
The positions of 22 newly discovered quasars (selected from 130 square
degrees) are indicated.  Already known quasars are not indicated in
this figure. 
}\label{colorobs}
\end{figure}

  Figure~\ref{colorobs} shows the color-color diagram of stellar
objects with $r^* < 20$ from 20 square degrees of SDSS imaging
commissioning data taken in September 1998\footnote{The asterisk $^*$
indicates that the final SDSS photometric system has not yet been
defined; this is preliminary photometry, accurate to perhaps 0.05
mag.}.  Notice the qualitative similarity to the simulations in
Figure~\ref{colorsim}, and the narrowness of the distribution: this is
a tribute both to the quality of the data, and the pipeline used to
reduce it.  As the SDSS spectrographs have not been commissioned as of
this writing, we are using the Double Imaging Spectrograph on the
Apache Point 3.5m telescope to carry out spectroscopy of promising
high-redshift quasar candidates.  Superposed on Figure~\ref{colorobs}
are the places in color-color space where the 22 new quasars we have
identified thus far lie, based on roughly 130 square degrees of
imaging data.


These quasars do not by any means constitute a complete sample.
In the last two nights of spectroscopic data, we have concentrated on
those objects which appeared from their broad-band colors to be
high-redshift candidates.  Out of 11 candidates, 9 are indeed quasars
at $z > 3.65$ (the two high-redshift quasars previously known in the
survey area also stood out cleanly in the color-color diagrams, and would
have been selected as well).  All are brighter than $i^* = 20$.  This
success rate far surpasses the typical 10\% found in the literature
for high-redshift quasar surveys\cite{Schneider,Hall,Kennefick}, 
although again,
we do not have a complete sample to make this quantitative. 

\begin{figure}
\centerline{\epsfig{file=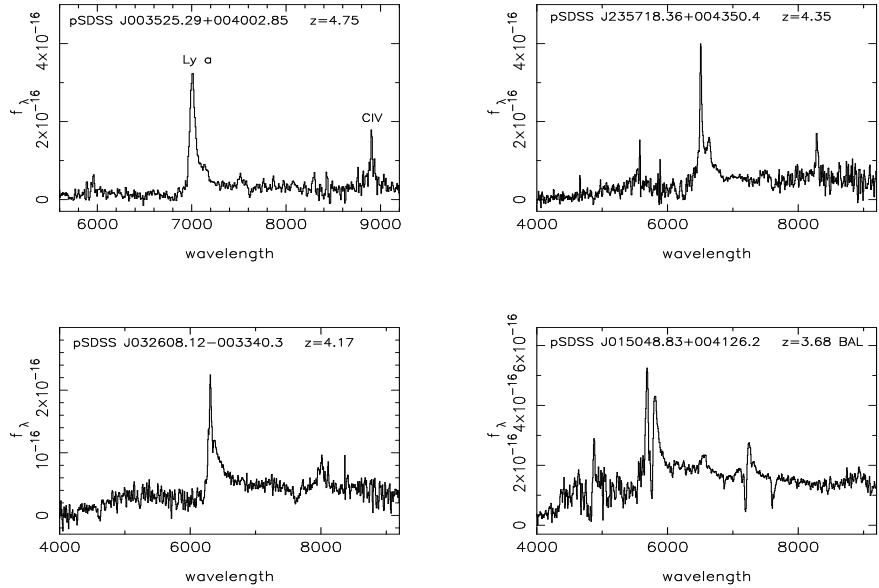,width=5in,height=2.5in}}
\vspace{10pt}
\caption{
Spectra of 3 new SDSS quasars with
$z>4$, plus one with a broad absorption-line spectrum, obtained with
the 3.5m ARC telescope and Double Imaging Spectrograph. 
}\label{qsospectra}
\end{figure}

Figure~\ref{qsospectra} shows our spectra of the three
highest-redshift quasars we have found thus far, plus one which shows
strong associated absorption.  The one at $z = 4.75$ is the
second-highest redshift quasar known (the current redshift holder is
$z = 4.90$; see ref.~\cite{Schneider91}).  These spectra are of quite
low resolution, roughly 7\AA\ pixel$^{-1}$, while the SDSS spectrographs will
deliver 1-1.5\AA\ pixel$^{-1}$ over a similar wavelength coverage.

These objects were selected from roughly 1\% of the sky that the SDSS
will image.  We therefore expect that there are enormously more 
high-redshift quasars to be discovered as part of the SDSS.

The Sloan Digital Sky Survey (SDSS) is a joint project of the
University of Chicago, Fermilab, the Institute for Advanced Study, the
Japan Participation Group, The Johns Hopkins University, Princeton
University, the United States Naval Observatory, and the University of
Washington.  Apache Point Observatory, site of the SDSS, is operated
by the Astrophysical Research Consortium.  Funding for the project has
been provided by the Alfred P. Sloan Foundation, the SDSS member
institutions, the National Science Foundation, NASA, and the U.S. Department
of Energy.  XF and MAS acknowledge additional support from Research
Corporation, NSF grant AST96-16901, the Princeton University
Research Board, and an Advisory Council Scholarship. 
DPS acknowledges support from NSF grant AST95-09919.


\begin{references}
\bibitem{GunnWeinberg}
Gunn, J. E., \& Weinberg, D. H. 1995, in {\it Wide-Field Spectroscopy
and the Distant Universe}, ed.\ Maddox and 
Arag\'on-Salamanca (Singapore: World Scientific), 3

\bibitem{SDSS} SDSS Collaboration, 1996, 
 {\tt http://www.astro.princeton.edu/BBOOK/}. 

\bibitem{Fukugita}  Fukugita, M., Ichikawa, T., Gunn, J.E., Doi, M., Shimasaku,
  K., \& Schneider, D.P. 1996, AJ, 111, 1748

\bibitem{CameraPaper} Gunn, J.E., Carr, M.A., Rockosi, C.M.,
Sekiguchi, M. {\sl et al.}~1998, AJ, in press 

\bibitem{Fan} Fan, X. 1998, AJ, submitted

\bibitem{Schneider} Schneider, D. P., Schmidt, M., \& Gunn, J.E. 1994,
AJ, 107, 1245

\bibitem{Hall} Hall, P.B., Osmer, P.S., Green, R.F., Porter, A.C., \&
Warren, S.J. 1996, AJ, 462, 614

\bibitem{Kennefick} Kennefick, J.D. {\sl et al.}~1995, 
AJ, 110, 78

\bibitem{Schneider91}  Schneider, D. P., Schmidt, M., \& Gunn, J.E. 1991,
AJ, 102, 837
\end{references}
\end{document}